\documentclass[apj]{emulateapj}

\pdfoutput=1

\usepackage{amsmath}
\usepackage{graphicx}
\usepackage{epstopdf}
\usepackage{epsfig}
\usepackage{natbib}
\usepackage{wasysym}
\usepackage{lipsum}
\usepackage{mathrsfs}
\usepackage{amsmath} % or simply amstext

\slugcomment{Accepted to the Astrophysical Journal Letters}

\usepackage{color}

\usepackage{filecontents}

\begin{document}

\shorttitle{Water Loss from Moist Greenhouse Atmospheres}
\shortauthors{Kasting et al.}

\title{Stratospheric Temperatures and Water Loss from Moist Greenhouse Atmospheres of Earth-like Planets}

\author{James F. Kasting\altaffilmark{1,2,3,4}}

\author{Howard Chen\altaffilmark{5,6}}

\author{Ravi K. Kopparapu\altaffilmark{1,2,3,4,7}}

\affil{\altaffilmark{1} Department of Geosciences, The Pennsylvania State University, State College, PA 16801, USA}
\affil{\altaffilmark{2} NASA Astrobiology Institute Virtual Planetary Laboratory}
\affil{\altaffilmark{3} Penn State Astrobiology Research Center, 2217 Earth and Engineering Sciences Building, University Park, PA 16802, USA}
\affil{\altaffilmark{4} Center for Exoplanets \& Habitable Worlds, The Pennsylvania State University, University Park, PA 16802, USA}
\affil{\altaffilmark{5} Department of Astronomy, Boston University, 725 Commonwealth Ave., Boston, MA 02215, USA}
\affil{\altaffilmark{6} Department of Physics, Boston University, 590 Commonwealth Ave., Boston, MA 02215, USA}
\affil{\altaffilmark{7} NASA Goddard Space Flight Center, 8800 Greenbelt Road, Greenbelt, MD 20771, USA}
\email{jfk4@psu.edu}

\begin{abstract}
A radiative-convective climate model is used to calculate stratospheric temperatures and water vapor concentrations for ozone-free atmospheres warmer than that of modern Earth. Cold, dry stratospheres are predicted at low surface temperatures, in agreement with recent 3-D calculations. However, at surface temperatures above 350 K, the stratosphere warms and water vapor becomes a major upper atmospheric constituent, allowing water to be lost by photodissociation and hydrogen escape. Hence, a moist greenhouse explanation for loss of water from Venus, or some exoplanet receiving a comparable amount of stellar radiation, remains a viable hypothesis. Temperatures in the upper parts of such atmospheres are well below those estimated for a gray atmosphere, and this factor should be taken into account when performing inverse climate calculations to determine habitable zone boundaries using 1-D models.
\end{abstract}

\keywords{astrobiology, planets and satellites: atmospheres, planets and satellites: terrestrial planets}

\section{Introduction} 
\label{sec:intro} 

The circumstellar habitable zone (HZ) is traditionally defined as the region around a star in which liquid water can remain stable on the surface of a rocky planet. According to standard theory \citep{KastingEt1993Icarus}, the inner edge of the habitable zone is set either by the onset of a runaway greenhouse, defined as complete evaporation of the oceans, or by the slightly earlier onset of a moist greenhouse, in which the stratosphere becomes wet and water is lost by photodissociation followed by hydrogen escape. This theory successfully explains the lack of water on our neighboring planet Venus, which formed somewhat inside the inner edge of the HZ.

In a recent paper, \citet{LeconteEt2013NATURE} used a 3-dimensional climate model to show that the runaway greenhouse threshold is pushed inward compared to 1-D calculations as a result of escape of longwave radiation through the undersaturated descending branches of the tropical Hadley cells. This result was welcome, as the most recent 1-D calculation \citep{KopparapuEt2013Erra,KopparapuEt2013ApJ} had placed this threshold at 0.99 astronomical units (AU), uncomfortably (and unrealistically) close to Earth's present orbit. The Leconte et al. paper moved it back to 0.95 AU, which is where it had been thought to lie for most of the past 37 years \citep{Hart1978Icarus,Kasting1988Icarus,KastingEt1993Icarus}.

\citet{LeconteEt2013NATURE} reached another conclusion, though, that challenges conventional thinking about how water might be lost from a Venus-like planet. As surface temperatures warmed from 280 K to 330 K in their model, stratospheric temperatures cooled from 140 K to below 120 K. (Following Leconte et al., we loosely refer to the atmospheric region above the troposphere as the stratosphere, although it could also be termed the mesosphere, as the ozone-free atmospheres being discussed lack the temperature inversion that is present in Earth's atmosphere.) This result was arguably not a numerical artifact, as the correlated-$k$ absorption coefficients in their model were derived for pressures as low as $10^{-6}$ bar. The low stratospheric temperatures, by themselves, are understandable, as the authors argued convincingly that such a result is to be expected if the atmosphere is distinctly non-gray (See also \citealt{Pierrehumbert2010}). The atmosphere modeled by Leconte et al. was highly non-gray because, along with 1 bar of N$_2$, it contained only 376 ppmv of CO$_2$. H$_2$O, while abundant near the surface, was almost completely absent from their model upper atmospheres. The stratosphere in their model is warmed by absorption of upwelling radiation in CO$_2$ line centers. Because these line centers are optically thick, they are shielded from the warm surface by CO$_2$ in lower atmospheric layers. Radiation to space can occur throughout the CO$_2$ absorption lines, though, and so the stratospheric temperature equilibrates at an extremely cold value.
	
The cold stratosphere in the Leconte et al. model appears to preclude the loss of water from a moist greenhouse planet, that is, one on which surface liquid water is still present. Indeed, the authors make this point explicitly in their paper. This result may not pose a problem in understanding water loss from Venus, as one recent study suggests that Venus never had liquid water on its surface \citep{HamanoEt2013NATURE}. Instead, Venus developed a true runaway greenhouse during accretion, and the steam atmosphere never condensed. Leconte et al. did not study runaway greenhouse atmospheres directly, but presumably such H$_2$O-rich atmospheres can always lose water by photodissociation and hydrogen escape \citep{Kasting+Pollack1983Icarus}.

This question may be relevant, though, to exoplanets near the inner edge of the circumstellar habitable zone. In older 1-D climate models (e.g. \citealt{Kasting1988Icarus,KastingEt1993Icarus}), the moist greenhouse occurs at a substantially lower stellar flux than a true runaway greenhouse. The more recent Kopparapu et al. (2013a,b) 1-D model does not show this large difference. All three of these studies employed inverse climate calculations in which the vertical temperature profile was specified, and radiative fluxes were back-calculated to determine the equivalent planet-star distance. These studies also all assumed an isothermal, 200 K stratosphere. This assumption was justified by comparison to a gray atmosphere model. In a gray atmosphere, the the temperature at optical depth zero, $T_{\rm 0}$ (also called the skin temperature) can be shown to be equal to the effective radiating temperature, $T_{\rm e}$, divided by $2^{1/4}$. For modern Earth, $T_{\rm e} \approx 255$ K, so $T_{0} \approx 214$ K. For Venus, $T_{\rm e} \approx 220$ K, so $T_0 \approx 185$ K. In the model of \citet{Kasting1988Icarus}, as the surface temperature warmed, the stratosphere became increasingly wet, allowing H$_2$O to be efficiently photodissociated and hydrogen to escape to space, even though liquid water was still present on Venus surface.
	
A new simulation of this problem of warm, moist planets performed with a different 3-D climate model (the NCAR CAM4 model) did find stable, moist greenhouse solutions \citep{Wolf+Toon2015}. Moist greenhouse solutions should be easier to achieve in 3-D models precisely because their tropospheres are undersaturated in some regions and because they include other climate feedbacks, such as clouds, that may help to stabilize a planet's climate. Leconte et al. (2013) did not find such solutions, but that is evidently not a general result. \citet{Wolf+Toon2015} found comparable stratospheric temperatures (always $\ga150$ K) to Leconte et al. at low surface temperatures, but at high surface temperatures they found much warmer (up to ${\sim}210$ K) stratospheric temperatures and correspondingly higher stratospheric H$_2$O mixing ratios \citep{Wolf+Toon2015}. Their absorption coefficients were derived for pressures down to $10^{-5}$ bar, so they should also be reliable in the upper stratosphere. (Their earlier paper \citet{Wolf+Toon2013} says that the lower pressure limit was only 0.01 bar, but this was evidently a typo, as evidenced by their accompanying discussion, and as confirmed by E. Wolf (priv. comm.).) If the Wolf \& Toon result is correct, then water would eventually be lost as a planet's surface temperature warms. %But which authors are correct, Leconte et al. or Wolf and Toon?    

\section{Model \& Approach}
\label{sec:model}
To answer this question, we used our own 1-D radiative-convective climate model, which has recently been updated to better handle runaway greenhouse atmospheres  \citep{KopparapuEt2013Erra,KopparapuEt2013ApJ}. Admittedly, our test is not definitive, because our 1-D model cannot simulate all of the processes included in a 3-D model. (In particular, although our model is non-gray, it cannot simulate the cold, high tropical tropopause which dries the stratosphere of modern Earth.) However, our model can predict vertical profiles of temperature and water vapor, and so it can be used as a sanity check on the 3-D results. The details of the model have been described in the reference just given, and so we will not repeat them here. One point deserves mention, though: The model uses correlated-$k$ absorption coefficients derived from the HITRAN and HITEMP databases for pressures of $10^{-5} - 10^2$ bar and for temperatures of 100-600 K. A pressure of $10^{-5}$ bar corresponds to an altitude of $\backsim 80$ km in the modern atmosphere. All calculations shown here use $10^{-5}$ bar as the pressure at the top of the model atmosphere. All calculations assume a noncondensable surface pressure of 1 bar of N$_2$, and most assume a CO$_2$ mixing ratio of 355 ppmv. Surface pressure increases as the temperature increases and H$_2$O becomes more abundant. O$_2$ and O$_3$ are excluded from the model.

Our 1-D model uses a time-stepping algorithm to reach steady-state solutions. Normally, we fix the solar flux and allow the model to compute a self-consistent vertical temperature/H$_2$O profiles. We term that the forward mode of calculation. Alternatively, the model can be run in inverse mode. In this case, we fix the surface temperature ($T_{\rm s}$), assume an isothermal stratosphere, and connect these to each other with a moist adiabat; then, we calculate the solar flux needed to sustain this surface temperature. Most, or all, of the runaway greenhouse calculations performed by Kasting (1988) and the more general HZ calculations performed by \citet{KastingEt1993Icarus} and  \citet{KopparapuEt2013ApJ,KopparapuEt2013Erra} were done in this manner. The reason is that runaway greenhouse atmospheres are$-$as their name implies$-$highly unstable.  As the solar flux is increased above its present value, $T_{\rm s}$ increases. This causes water vapor to increase, which causes $T_{\rm s}$ to increase further, until eventually the model ``runs away" to very high surface temperatures. Indeed, with our current set of H$_2$O absorption coefficients, which are derived from the HITEMP database, our model runs away at the Earth's current solar flux if the troposphere is assumed to be fully saturated. A saturated troposphere is not realistic for the modern Earth, but it becomes a better and better approximation as the atmosphere becomes warmer and more water-rich. Treating relative humidity self-consistently requires a 3-D model like the ones developed by \citet{LeconteEt2013NATURE} and \citet{Wolf+Toon2015}.

Inverse calculations are stable and easy to perform with our 1-D model; however, they require that the stratosphere be isothermal, which is precisely the assumption that has been challenged by Leconte et al. So, we modified our 1-D model to do a type of calculation that is somewhere in between the forward and inverse modes. We used a time-stepping procedure, as in the forward model; however, after each time step we reset the surface temperature to a specified value. We then determined where the atmospheric cold trap is located, typically somewhere in the lower stratosphere, and we reset temperatures below that level (but not including the surface) to the cold trap temperature. The cold trap is the altitude at which the saturation mixing ratio of H$_2$O is at a minimum, so it determines the stratospheric H$_2$O concentration. (Note that this is not necessarily the altitude at which the stratospheric temperature is lowest. If a low temperature occurs at a correspondingly low pressure, p, then the saturation mixing ratio of water vapor, p$_{\rm sat}$/p, may still be relatively high.) We next drew a moist adiabat up from the surface until it intersected the temperature profile that had been formed in that way. We then recomputed fluxes and repeated the entire procedure until the temperature profile reached steady state. This methodology allowed upper stratospheric temperatures to achieve radiative equilibrium while preventing the surface temperature from running away.

\section{Results \& Discussion}
\label{sec:results}
Calculations were performed for various surface temperatures ranging from 288 K (the present value for Earth) up to 370 K. Results are shown in Fig. 1. Fig. 1a shows temperature profiles. At low $T_{\rm s}$, our model predicts upper stratospheric temperatures of 100 K, or even lower$-$well below the temperatures predicted by either of the 3-D climate models discussed earlier. But at higher surface temperatures, the stratospheric temperature rises, as it does in the \citet{Wolf+Toon2015} model. For $T_{\rm s} = 370$ K, the temperature at the top of the convective troposphere is ${\sim}200$ K, right where it was assumed to be in the \citet{Kasting1988Icarus} model. Admittedly, the stratosphere is not well resolved in this particular calculation, as nearly the entire atmosphere is convective by this point. We did not attempt to extend the model higher, though, because our absorption coefficients are only good down to $10^{-5}$ bar.

 Qualitatively, these temperature profiles look much like those in Kasting (1988, Fig. 5a) except that the stratosphere is no longer isothermal. At low temperatures the convective layer extends up to only ${\sim} 12$ km, but at high temperatures it extends well above 100 km. This dramatic difference is caused by the increased importance of latent heat release, which causes the lapse rate to become shallower at high surface temperatures.

\begin{figure}[t] %different options for where to place figure
\begin{center}
\includegraphics[width=0.8\columnwidth]{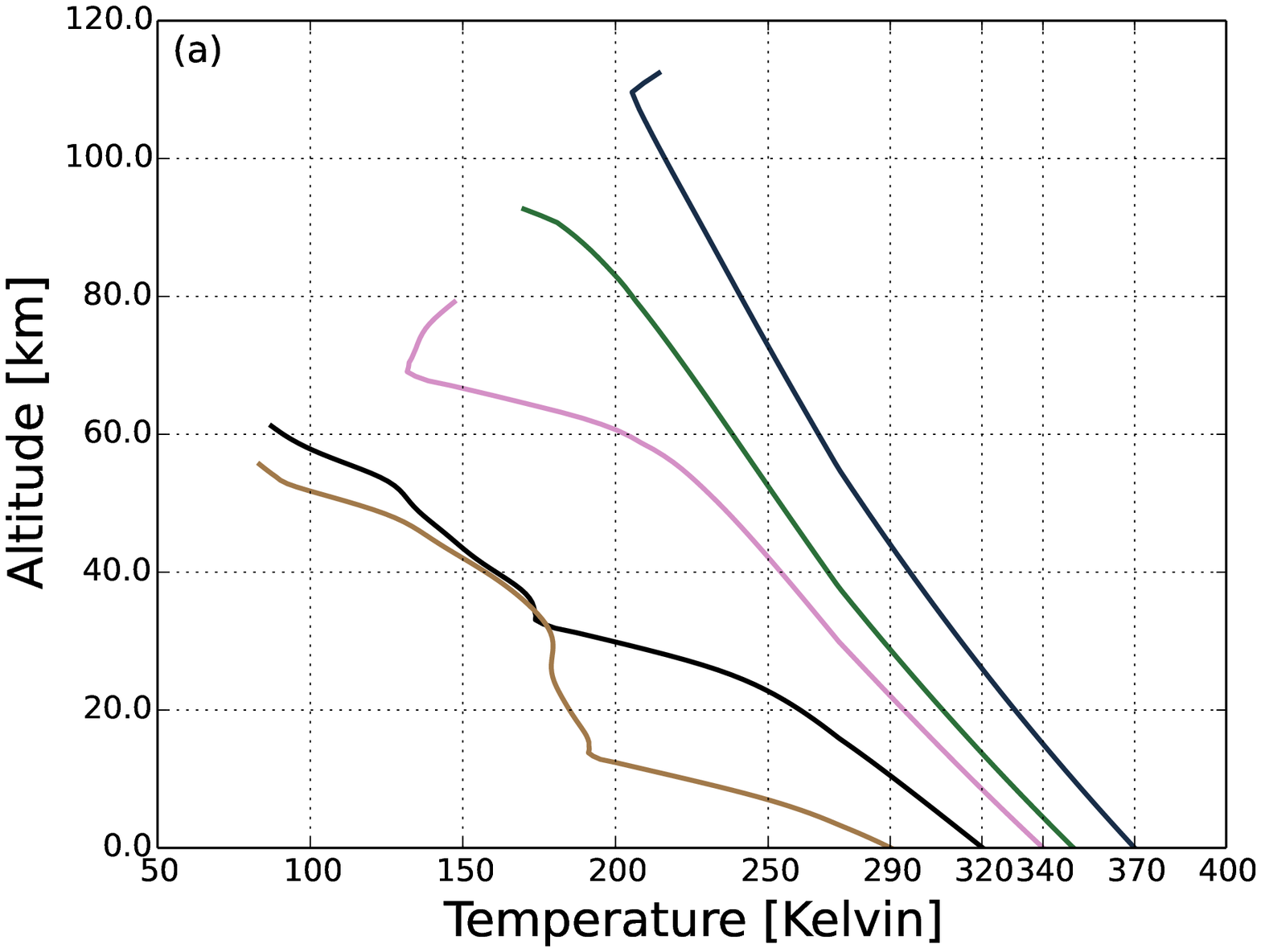}
\includegraphics[width=0.8\columnwidth]{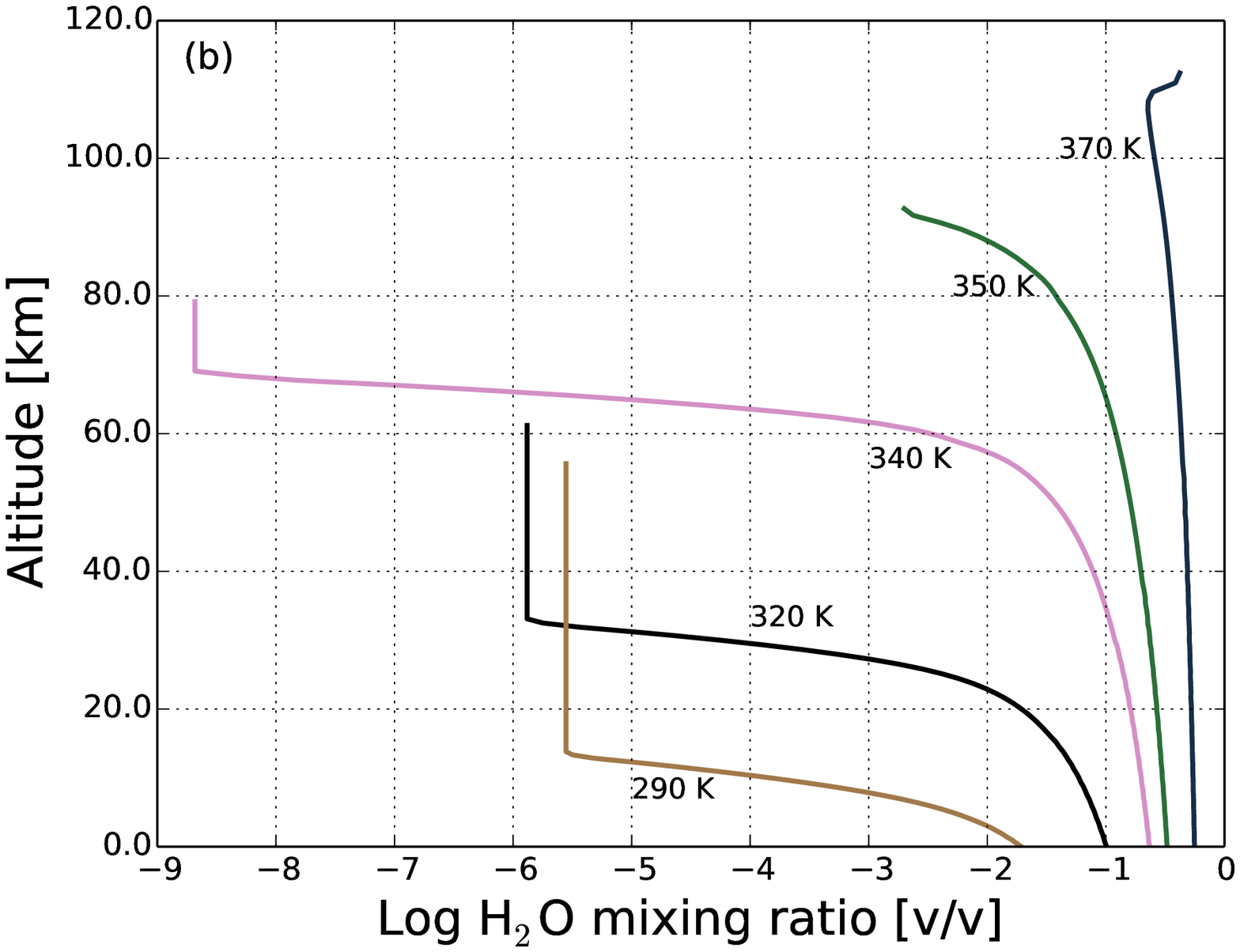}
\caption{\label{fig:leconte}
Vertical profiles of temperature (panel a) and water vapor (panel b) calculated using our 1-D radiative-convective climate model. The assumed CO$_2$ concentration is 355 ppmv. }
\end{center}
\end{figure}

\begin{figure}[t] %different options for where to place figure
\begin{center}
\includegraphics[width=0.8\columnwidth]{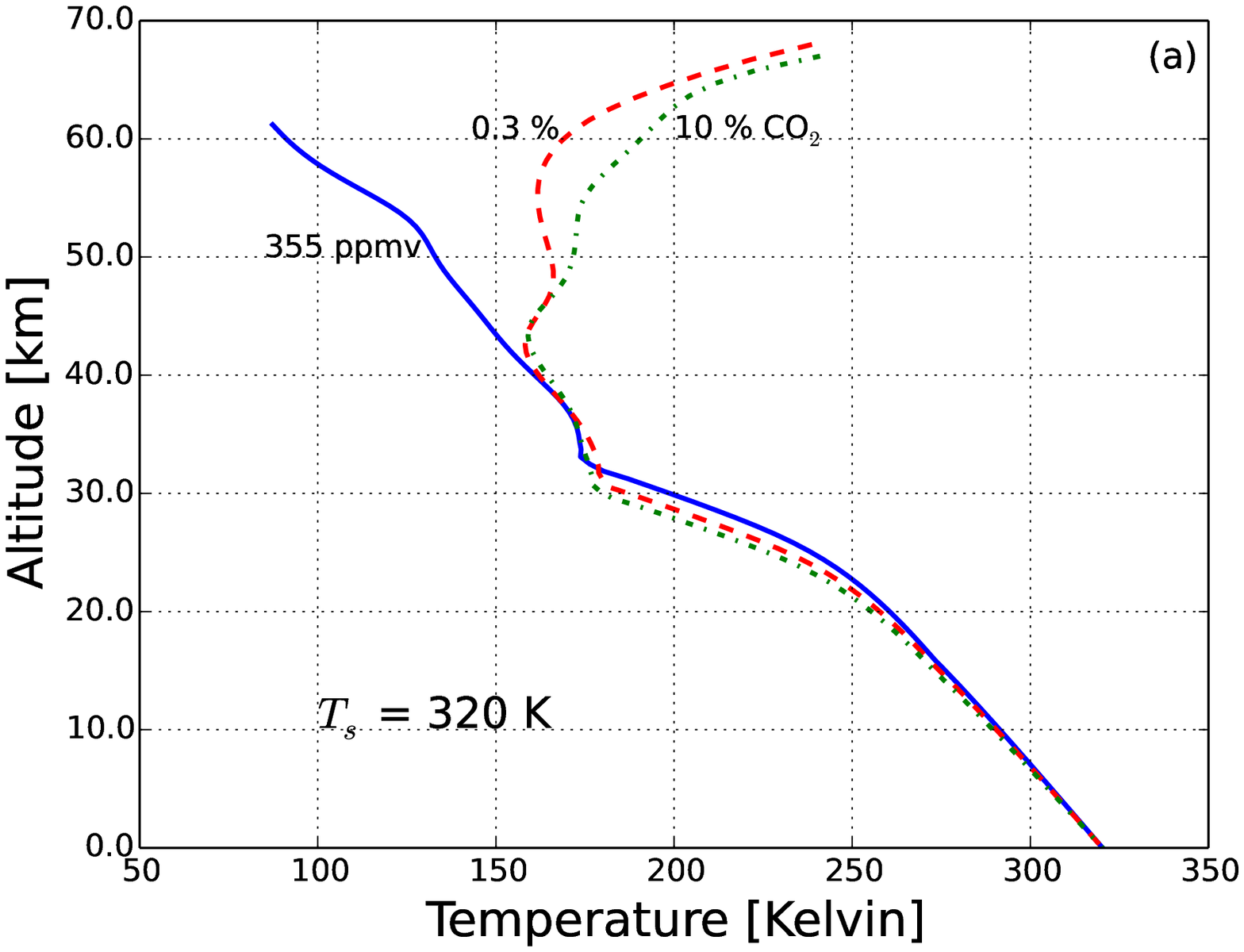}
\includegraphics[width=0.8\columnwidth]{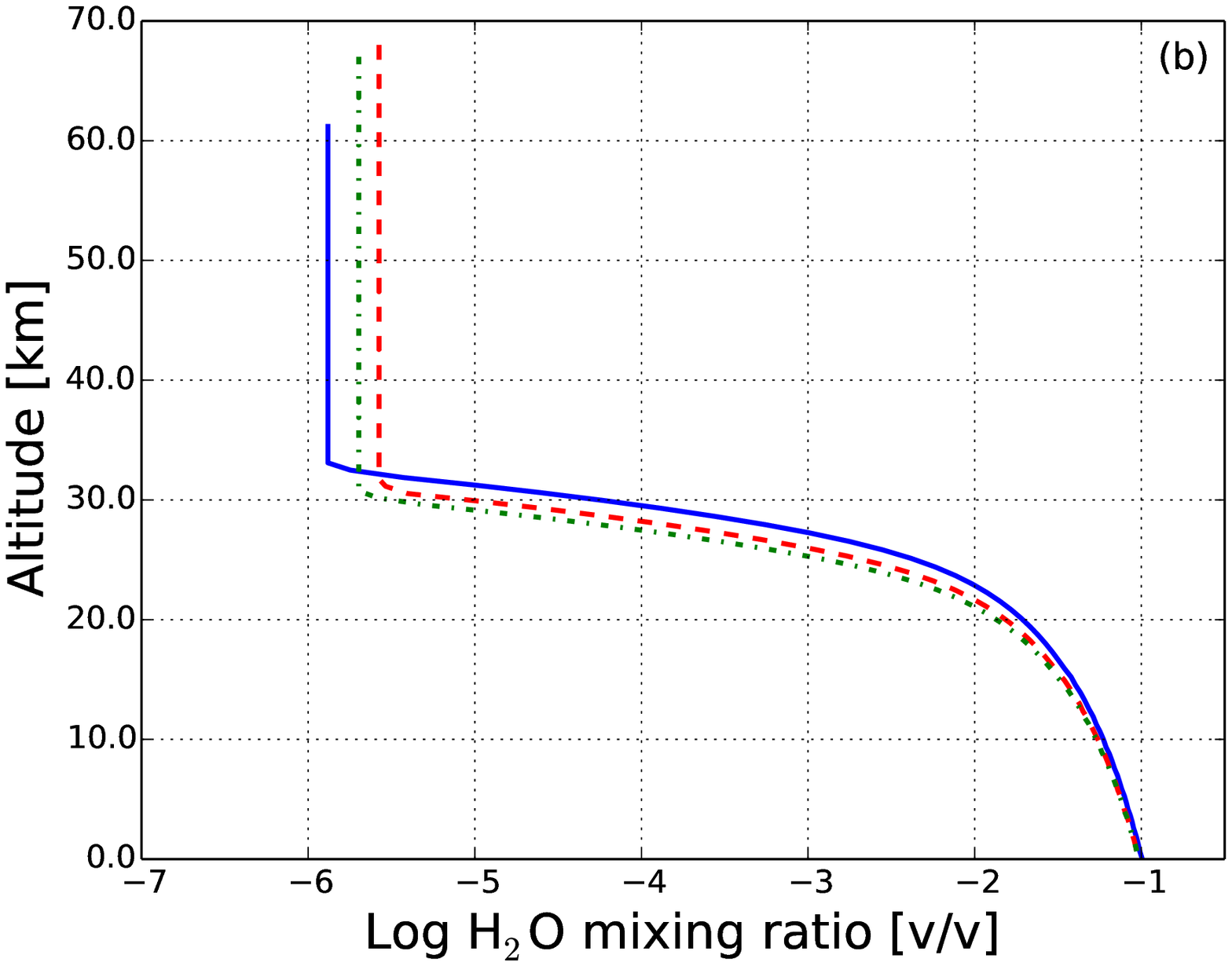}
\caption{\label{fig:increaseco2}
Vertical profiles of temperature  (panel a) and water vapor (panel b) for different atmospheric CO$_2$ concentrations. The assumed surface temperature is 320 K. Note that the amount of water vapor does not experience significant change in response to the rise in CO$_2$. }
\end{center}
\end{figure}

Corresponding water vapor profiles are shown in Fig. 1b. At first glance, these profiles again look much like those in Kasting (1988, Fig. 5b). At low surface temperatures, water vapor is a minor constituent of the stratosphere, as it is in Earth's atmosphere today. At high surface temperatures, water vapor becomes a major atmospheric constituent at all altitudes. Because the hydrogen escape rate is proportional to the total hydrogen mixing ratio in the upper atmosphere \citep{Kasting+Catling2003}, this means that water could readily be lost from our high-$T_{\rm s}$ atmospheres. 

If one looks more closely, however, significant differences from the \citet{Kasting1988Icarus} calculations can be seen at intermediate values of surface temperature. At $T_{\rm s} = 320$ K, the older model predicted a stratospheric H$_2$O mixing ratio of nearly $10^{-4}$, whereas the current model predicts a value closer to $10^{-6}$. And, at $T_{\rm s} = 340$ K, the discrepancy is even larger: the older model predicted a stratospheric H$_2$O mixing ratio of ${\sim}10^{-3}$, whereas the new model predicts a value of ${\sim}10^{-9}$. These differences are caused by the much colder stratospheric temperatures in the present model. But, as $T_{\rm s}$ increases further and stratospheric H$_2$O becomes more abundant, stratospheric temperatures increase, as well. This behavior can be physically explained: H$_2$O absorbs well across much of the thermal-infrared spectrum; so, as H$_2$O becomes more abundant, the atmosphere becomes more and more like a gray atmosphere. We have already seen that, given Earth-like insolation, the skin temperature of a gray atmosphere should be in the neighborhood of 200 K. Our calculated stratospheric temperatures tend towards that value as H$_2$O becomes abundant.
	
It is easy to see why the Leconte et al. model does not exhibit this behavior. The highest surface temperature reached in their calculation is only 330 K. At this point in our own calculations, the stratosphere is still cold and dry. But when $T_{\rm s}$ reaches 350 K, water vapor begins to break through into the stratosphere, and the stratosphere begins to warm. We are able to explore this temperature regime because of the ease with which one can manipulate a 1-D model. But it is more difficult to do this in 3-D climate model because such models must always be run in forward, time-stepping mode and because the included physical parameterizations (e.g., moist convective fluxes) are often more complex.

As mentioned previously, \citet{Wolf+Toon2015} do obtain solutions up to surface temperatures of ${\sim}370$ K, well above the 330 K reached by Leconte et al. Their model exhibits negative cloud feedback at high surface temperatures, which helps stabilize the climate in this regime. Their model, like ours, predicts that stratospheric water vapor increases smoothly as $T_{\rm s}$ increases. At low $T_{\rm s}$, their calculated stratospheric temperatures are also consistently warmer than either ours or those of \citet{LeconteEt2013NATURE}, for reasons that are unclear. At $T_{\rm s}$ = 370 K, their stratospheric temperature is ${\sim}210$ K, just like ours. Wolf \& Toon computed absorption coefficients at 56 different pressure levels, as compared to 8 levels in our model and 9 in the Leconte et al. model, so it is possible that their finer pressure resolution results in increased accuracy. But their model also develops a temperature inversion near the surface, which may be unphysical. (How does the surface remain in thermal balance when convection is absent and the temperature is higher both above and below the surface?) So, it is still worth investigating this question with an independent model.
	
We performed one further set of calculations to explore the dependence of these results on the atmospheric CO$_2$ concentration. With $T_{\rm s}$ fixed at 320 K, we increased the CO$_2$ mixing ratio to 0.3\% and 10\%. Results are shown in Fig. 2. Surprisingly (or perhaps not), the stratospheric temperature warms as the CO$_2$ concentration is increased. This result may seem surprising at first, as CO$_2$ is regarded as a coolant in Earth's modern stratosphere. But the modern stratosphere is relatively warm because of absorption of solar UV radiation by ozone. In the extremely cold stratospheres modeled here, the only significant heating comes from absorption of upwelling thermal-IR radiation by CO$_2$; hence, adding more CO$_2$ has a warming effect. Conversely, increasing CO$_2$ had little effect on stratospheric H$_2$O concentrations (Fig. 2b).

One further observation can be made based on these results. One can still calculate a moist greenhouse limit using a 1-D climate model, but one needs to be careful in doing so, as the assumption of an isothermal 200 K stratosphere is clearly invalid. If one has to pick a stratospheric temperature, 150 K would be a better estimate for a low-CO$_2$ atmosphere. The moist greenhouse limit, like the runaway greenhouse limit, should ideally be calculated using 3-D climate models.\\

\section{Conclusion}
\label{sec:conclude}
Our calculations support the results of \citet{LeconteEt2013NATURE} in that we, too, find very low stratospheric temperatures for moderately warm, low-CO$_2$ atmospheres that lack O$_2$ and O$_3$. And we, too, calculate low stratospheric H$_2$O concentrations for surface temperatures up to ${\sim}340$ K. At still higher surface temperatures, however, the stratosphere warms and H$_2$O becomes a major upper atmospheric constituent, as in the earlier model of \citet{Kasting1988Icarus} and the more recent model of \citet{Wolf+Toon2015}. Thus, contrary to the claim of \citet{LeconteEt2013NATURE}, water loss does appear to be possible from a moist greenhouse planet. Finally, our calculations suggest that the moist greenhouse limit for the inner edge of the habitable zone can be estimated by doing 1-D inverse calculations, provided that one uses a stratospheric temperature of 150 K, instead of the canonical value of 200 K used in earlier studies. But a fully saturated 1-D climate model will likely underestimate the solar flux needed to trigger a moist greenhouse and will thus produce a habitable zone inner edge that is too far away from the star. More accurate estimates of the inner edge boundary require the use of 3-D climate models.

\acknowledgements
The authors thank the anonymous referee for insightful comments. H.C. acknowledges the Undergraduate Research Opportunities Program (UROP) at Boston University for primarily funding the research while in residence at Penn State University in the summer of 2015. J.F.K. and R.K. thank NASA's Emerging Worlds and Exobiology programs for their financial support.

%\bibliography{earthplanets}

%\bibliographystyle{apj}

\end{document}